\documentclass{Interspeech2023}

\title{
    audb - Sharing and Versioning of Audio and Annotation Data in Python
}

\name{
Hagen Wierstorf\,\textsuperscript{1},
Johannes Wagner\,\textsuperscript{1},
Florian Eyben\,\textsuperscript{1},
Felix Burkhardt\,\textsuperscript{1},
Bj\"{o}rn W. Schuller\,\textsuperscript{1,2,3}
}
\address{
    \textsuperscript{1} audEERING GmbH, Gilching, Germany\\
    \textsuperscript{2} Chair of Embedded Intelligence for Health Care and Wellbeing, University of Augsburg, Germany\\
    \textsuperscript{3} GLAM -- Group on Language, Audio, \& Music, Imperial College, UK
}
\email{hwierstorf@audeering.com}
\interspeechcameraready

\usepackage{listings}
\usepackage{xcolor}
\usepackage[scaled]{DejaVuSansMono}
\usepackage[T1]{fontenc}
\definecolor{codegreen}{HTML}{598006}
\definecolor{codeblue}{HTML}{005295}
\definecolor{codegray}{HTML}{6a747c}
\definecolor{codeorange}{HTML}{bb5101}
\definecolor{backcolour}{rgb}{1,1,1}
\lstdefinestyle{mystyle}{
    backgroundcolor=\color{backcolour},   
    commentstyle=\color{codegray},
    keywordstyle=\color{codeblue},
    numberstyle=\tiny\color{codeorange},
    stringstyle=\color{codegreen},
    basicstyle=\ttfamily\footnotesize,
    breakatwhitespace=false,         
    breaklines=true,                 
    captionpos=b,                    
    keepspaces=true,                    
    numbersep=5pt,                  
    showspaces=false,                
    showstringspaces=false,
    showtabs=false,                  
    tabsize=4
}
\usepackage[natbib, bibencoding=utf8, citestyle=numeric, bibstyle=ieee, maxbibnames=999, maxcitenames=2, mincitenames=1, sortcites]{biblatex}
\bibliography{bibliography}

\usepackage[capitalise, nameinlink, noabbrev]{cleveref}

\usepackage[acronym, shortcuts, nohypertypes={acronym}]{glossaries}
\newacronym{CCC}{CCC}{concordance correlation coefficient}
\newacronym{SER}{SER}{speech emotion recognition}
\newacronym{PCC}{PCC}{Pearson correlation coefficient}
\newacronym{MIR}{MIR}{Music Information Retrieval}

\begin{document}

\maketitle

\begin{abstract}
Driven by the need for larger and more diverse datasets
to pre-train and fine-tune increasingly complex machine learning \mbox{models},
the number of datasets is rapidly growing.
audb is an open-source Python library
that supports versioning and documentation of audio datasets.
It aims to provide a standardised and simple user-interface
to publish, maintain, and access the annotations and audio files
of a dataset.
To efficiently store the data on a server,
audb automatically resolves dependencies 
between versions of a dataset
and only uploads newly added or altered files
when a new version is published.
The library supports partial loading of a dataset
and local caching for fast access.
audb is a lightweight library
and can be interfaced from any machine learning library.
It supports the management of datasets
on a single PC,
within a university or company,
or within a whole research community.
\end{abstract}

\glsresetall

\section{Introduction}
\label{sec:introduction}

To foster progress in automatic speech emotion recognition
and related learning tasks,
it is crucial to have a quick and easy way
of accessing an ensemble of datasets
for training and evaluation~\citep{scheidwasser-clow2022}.
This requires that the datasets
have a unique identifier,
are versioned,
can be shared and combined,
are documented in a standardised way~\citep{gebru2021datasheets},
and can be accessed from a common user interface.


This paper presents \emph{audb},
a Python library to publish, maintain, and access
labelled or unlabelled audio data in machine learning pipelines.
It also supports audio tracks embedded in video files.
It can load a dataset by name and version from different repositories.
The dataset is then provided in a well defined specification
(\emph{audformat}\footnote{https://audeering.github.io/audformat/}),
and its audio data can be resampled,
remixed,
or converted to the desired format.
A caching mechanism guarantees quick access.
A dataset consists of a root folder with a header file
and multiple table files
holding metadata and annotations,
and the referenced audio files,
usually organised into sub-directories.
The Python library
\emph{audinterface}\footnote{https://audeering.github.io/audinterface/}
provides an interface to read and process
the audio data of one or more datasets.

\emph{audb} is under continuous development
and has been used
to publish and maintain 840 datasets and versions
since 4 years inside audEERING.
It is released open-source since 2021 under an MIT license,
available via PyPI\footnote{https://pypi.org/project/audb/}
and Github,\footnote{https://github.com/audeering/audb/}
and the documentation is hosted on the project website.\footnote{https://audeering.github.io/audb/}

\section{Related Work}

With the introduction of Git in 2005
and platforms like Github in 2008
for development and sharing of code, 
it became obvious that no convenient solution
for audio data management and sharing
existed within the research community.
In 2014,  
Git Large File Storage\footnote{https://git-lfs.com}
was released and established
as the standard way of including binary files
in git repositories
based on similar ideas like
git-media\footnote{https://github.com/alebedev/git-media}
which existed already since 2009.
With this approach, it became possible
to have git repositories
that provide versioning of data
and track authorship of certain changes to the data.
As Git Large File Storage
did not focus on a particular kind of binary data to be versioned
Data Version Control\footnote{https://dvc.org}
evolved since 2017
with a focus on versioning data,
machine learning models,
and experiments
to foster reproducibility~\citep{stodden2010reproducible,olorisade2017reproducibility, pawlik2019}.

In parallel, the problem of sharing large amount of research data
was tackled by approaches like Zenodo
established in 2013~\citep{purcell2013zenodo}.
Zenodo allows researchers to upload datasets
and provides a digital object identifier~\citep{iso2022doi}
to make datasets easier to cite
and provide long-term access to them~\citep{herterich2016data}.
Access to shared data can be improved
if the data and its corresponding metadata or annotations
are also provided in a standardised way.
One successful example from the audio community is
the Spatially Oriented Format for Acoustics (SOFA) format
for impulse responses~\citep{majdak2013}.

Recently,
different audio communities have addressed the problem of reproducibility
with open-source toolkits,
which help to re-run experiments and access related datasets.
\citet{bittner2019mirdata} introduced a Python library
to load and manage annotations for \ac{MIR} datasets,
which was later extended for more general audio datasets~\citep{fuentes2021soundata}.
The audio source separation community develops the Asteroid toolkit
which can access relevant datasets~\citep{pariente2020asteroid}.
More general toolkits like
SpeechBrain~\citep{ravanelli2021speechbrain},
TensorFlow~\citep{abadi2016tensorflow}, or 
PyTorch~\citep{paszke2019pytorch}
include handling of data,
but do not focus on data versioning and management.
The Hugging Face \emph{Datasets}~\citep{lhoest2021datasets} library
extends the dataset handling from TensorFlow
and makes it independent of any machine learning library.
It provides access to datasets
for natural language processing,
but also computer vision, and audio.
\emph{Datasets} can efficiently handle very huge datasets
by streaming the data
and loading it only partially into memory.
On the Hugging Face Hub\footnote{https://huggingface.co/datasets},
it provides repositories for datasets
in which the versioning is handled by Git
and Git Large File Storage.
As \emph{Datasets} addresses data management
in a similar way to \emph{audb},
\cref{sec:huggingface} will compare them in more depth.

\section{Library Overview and Design}

The most important functionality of the library
is to load a dataset and access its annotations
and files.
The following example loads 
version 1.3.0 of the emodb dataset~\citep{burkhardt2005emodb}
and returns the file names and corresponding annotations
stored in a table with the name `emotion'
as a pandas dataframe:

\begin{lstlisting}[language=Python]
db = audb.load("emodb", version="1.3.0")
df = db["emotion"].get()
\end{lstlisting}

\noindent
A complete list of all available functions and classes with examples
is provided with the \emph{audb} API documentation.\footnote{https://audeering.github.io/audb/api/audb.html}

\subsection{Annotations, Metadata and Header}

Annotations are stored as columns in tables,
which are represented in a human readable way by CSV files
named `db.<table id>.csv',
and cached as pickle files for faster access.
Each table and column is identified by a unique ID.
The rows in the tables are associated
with audio files or segments of audio files,
which define the index of the table.
A \emph{filewise} index is used
to reference files as a whole:

\begin{lstlisting}
file,emotion
a.wav,happy
b.wav,angry
\end{lstlisting}

\noindent
Or if segments should be referenced,
a \emph{segmented} index
with additional start and end times is used:

\begin{lstlisting}
file,start,endemotion
c.wav,0 days 00:00:01.0,0 days 00:00:03.3,happy
c.wav,0 days 00:00:03.5,0 days 00:00:07.8,angry
\end{lstlisting}

\noindent
Annotations that are not attached to a file
can be organised in \emph{misc tables},
which support custom indices.
e.\,g., \,the following table stores 
age and gender of the speakers in the dataset:

\begin{lstlisting}
speaker,age,gender
spk0,29,female
spk1,93,male
\end{lstlisting}

\noindent
It is possible to restrict the values in a column
of a (misc) table to a certain data type or range
(bool, date, float, integer, object, string, time)
by assigning it to a scheme.
For example,
a column with annotations of emotion
can be restricted to a set of labels like
`happy`, `angry`, `neutral`.
This information is stored in the header of a dataset.

\begin{table}[t]
    \centering
    \caption{
        Default metadata entries in the header of a dataset.
        If needed, 
        the list can be extended by custom fields.}
    \begin{tabular}{lll}
        \toprule
        \textbf{Field} & \textbf{Mandatory} & \textbf{Description} \\
        \midrule
        name           & yes & name of dataset \\
        source         & yes & original source, e.\,g.,  URL \\
        usage          & yes & data usage, e.\,g.,  research \\
        author         &     & author(s) \\
        description    &     & long description \\
        expires        &     & expiration date if applicable \\
        languages      &     & included languages \\
        license        &     & license \\
        organisation   &     & organisation \\
        \bottomrule
    \end{tabular}
    \label{tab:metadata}
\end{table}

In addition, 
the header lists information about
attachments, tables, columns, raters, and splits.
And it stores metadata about the dataset,
which is summarised \cref{tab:metadata}.
The header is saved in a YAML file, 
which is located in the root folder of the dataset.

\subsection{Repositories and Backends}

\emph{audb} can host datasets in one or more repositories,
which can be distributed over different backends.
Currently,
\emph{audb} supports a local file system,
and an Artifactory instance
as backend.
But it is also possible to implement custom backends
and register them with \emph{audb}.

The repository of a dataset is defined by 
a name, a host, and a backend,
and can be obtained via:

\begin{lstlisting}[language=Python]
repo = audb.repository("emodb", version="1.3.0")
repo.name
repo.host
repo.backend
\end{lstlisting}

\subsection{Publication and Versioning}
\label{sec:publishing}

A new dataset is published from a root folder,
which contains its header and tables,
as well as,
the referenced audio files,
which are possibly organised into sub-folders.
For instance,
consider a dataset with two audio files
and a table with ID `emotion',
stored in a folder `dataset':

\begin{lstlisting}
dataset/
    audio/
        a.wav
        b.wav
    db.emotion.csv
    db.yaml
\end{lstlisting}

\noindent
The dataset can then be published
as version 1.0.0
to some repository with:

\begin{lstlisting}[language=Python]
audb.publish("./dataset", "1.0.0", repository)
\end{lstlisting}

\noindent
The dataset header, table, and audio files
are uploaded as individual ZIP files to the repository.
In addition,
a dependency table is created,
which holds metadata about audio files
(e.\,g., sampling rate)
and records for every table and audio file
in which version of the dataset it is stored.
The entries in the dependency table are also available to the user,
e.\,g.,  compare the sampling rate entry in the model card in \cref{fig:emodb}.

\begin{figure}[t]
    \centering
    \includegraphics[width=\columnwidth]{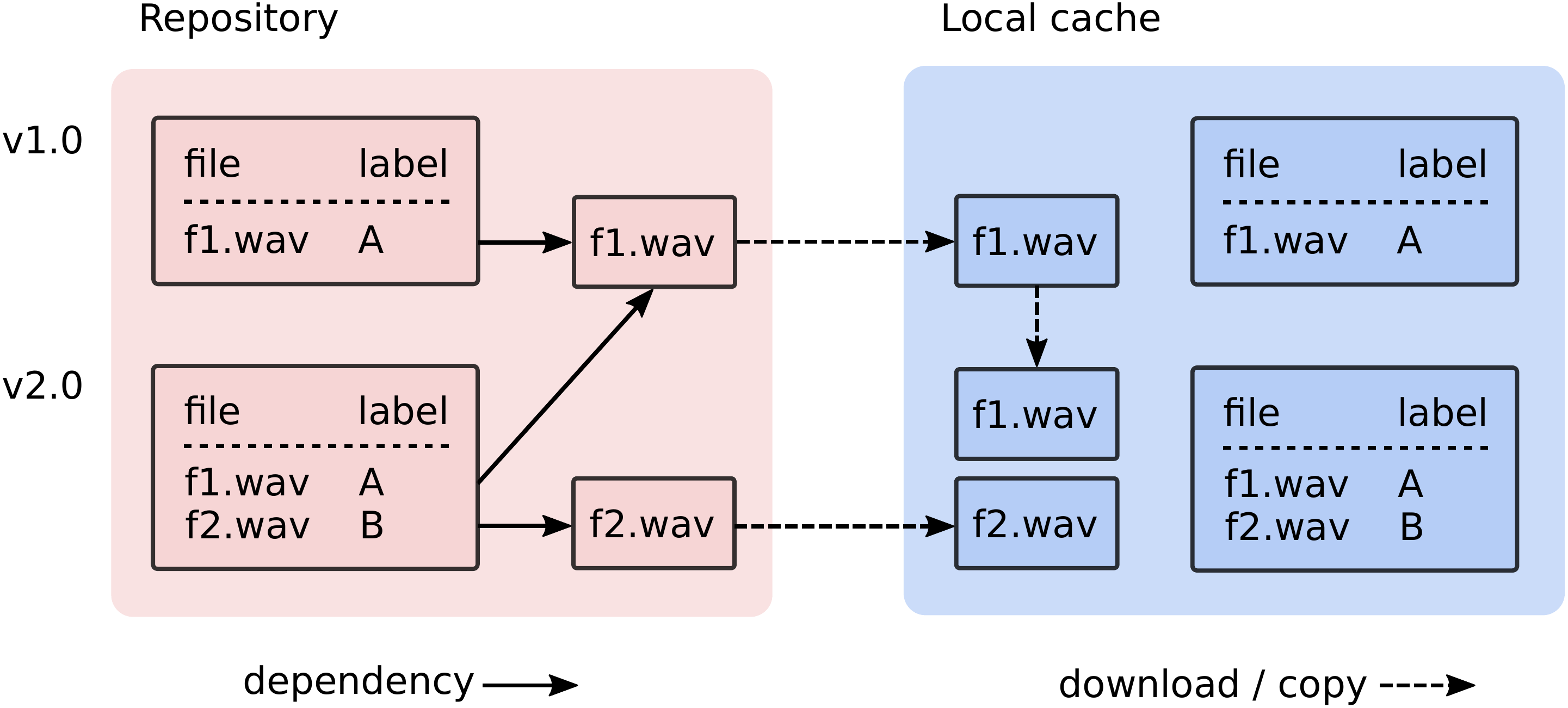}
    \caption{
        A dependency system ensures
        that only new or altered files are uploaded
        to the server for new dataset versions.
        For instance,
        in v2.0, a dependency to `f1.wav` from v1.0 is set
        (left part).
        When the dataset is loaded,
        references are resolved
        and a self-contained copy of the dataset is created
        in the cache.
        If possible, files are retrieved
        from other versions of the dataset
        that exist in the cache.
        For instance,
        when loading v2.0 the file `f1.wav` is copied from v1.0
        (right part).
    }
    \label{fig:repo-cache}
\end{figure}

To publish a new version of a dataset,
a user downloads header and tables
of a previous version,
and optionally also audio files
if she plans to replace them.
Now,
existing files can be deleted or modified,
and new files can be added.
Afterwards,
the dataset can be published 
under a new version.
During publication
\emph{audb} automatically identifies the changes
and uploads only the new or altered files.
For the remaining files, a dependency to the version
in which it was last modified is set
(see \cref{fig:repo-cache}).

For well established datasets
or datasets that grow over time, 
we recommend to version the scripts
that publish the dataset
on a service like Github.
This allows users to open issues
or create pull requests to fix errors in the dataset.
An example of such a dataset project
for the emodb dataset~\citep{burkhardt2005emodb}
can be found at \url{https://github.com/audeering/emodb}.

Since \emph{audb} handles the versioning
and can automatically detect changes made to a dataset,
it is possible to fully automate the publishing process.
This allows it to directly integrate the data publishing
into annotation or data collection tools.

\subsection{Flavours}

When loading a dataset,
by default,
the original audio files are retrieved.
However,
\emph{audb} offers the option
to request a dataset in a specific flavour.
In that case
the audio files are converted to the same format
with a specific bit depth,
sampling rate,
channel selection,
and mix-down.
For each flavour,
a separate cache folder is used,
i.\,e., the same dataset may be available in different formats.
In a machine learning pipeline,
flavours can be used to ensure that 
audio files stemming from different datasets 
are in the same format,
e.\,g., \,share a sampling rate of $8000$\,Hz:

\begin{lstlisting}[language=Python]
db = audb.load("emodb", sampling_rate=8000)
\end{lstlisting}

\subsection{Partial Loading}
\label{sec:partial-loading}

To speed up loading,
it is possible to request only specific parts of a dataset.
For example,
the header and tables of a dataset
can be loaded without audio files:

\begin{lstlisting}[language=Python]
db = audb.load("emodb", only_metadata=True)
\end{lstlisting}

\noindent
Or specific tables can be loaded,
which will only load audio files
referenced in those tables:

\begin{lstlisting}[language=Python]
db = audb.load("emodb", tables="emotion")
\end{lstlisting}

\noindent Or specific audio files can be loaded,
which will automatically remove
other entries from the tables:

\begin{lstlisting}[language=Python]
db = audb.load("emodb", media="wav/03a01Fa.wav")
\end{lstlisting}

\subsection{Caching}

When a dataset is loaded,
\emph{audb} figures out missing tables and audio files,
and either copies them from an already cached version of the dataset 
or, if that is not possible, downloads them from the server
(see \cref{fig:repo-cache}).
If the dataset is completely cached,
loading works without an internet connection.
Inside the cache a folder is created
for every version and flavour of a dataset.
Dependencies to earlier versions are automatically resolved
so that the folder in which the dataset is stored
contains all files.
This consumes more space,
but has the advantage
that the dataset is self-contained
and can be shipped as is and directly loaded with \emph{audformat}.
Tables are cached as CSV files,
and in addition pickled for fast reading.

\subsection{Removing Audio Files From All Versions}

Audio recordings may contain sensitive information.
Therefore, \emph{audb} offers the option
to remove specific audio files
from all published versions.
This goes beyond dropping files
with a new dataset version
as discussed in \cref{sec:publishing},
which does not remove files
from previous versions.
This can result in non-reproducibility of some results,
but avoids completely removing affected versions of the dataset.

\section{Comparison with Hugging Face Datasets}
\label{sec:huggingface}

The Hugging Face Hub
provides data repositories to publish datasets
with Hugging Face \emph{Datasets}. 
A data repository contains documentation of the dataset
and a Git repository with Git Large File Storage support
to version the data.
A data repository can contain
a so called loading script,
which \emph{Datasets} executes when loading the data.
This allows \emph{Datasets} to download data from external sources
and easily incorporate public datasets
already stored somewhere, e.\,g.,  on Zenodo.
As a consequence of this approach, 
\emph{Datasets} lacks information
which files persist between versions
of a dataset
and therefore all data (again) has to be downloaded
when a new version of a dataset is requested.
In contrast,
\emph{audb} does not support linking external sources
as all audio files must be part of the repository.
This,
however,
enables \emph{audb} to store and load the data more efficiently
since the same file can be shared across versions.
Another disadvantage of loading datasets
with a script
is that rolling out a dataset can be slow,
as it might require parsing a million lines of annotations first
and convert them from row to column representation.

\emph{audb} can download
single audio files from a dataset,
whereas with \emph{Datasets}, 
this is only possible if the creator of a dataset
puts each audio file into a single archive
with the name of the audio file
so that it can be addressed during download.
However,
the common approach with \emph{Datasets}
is to not publish individual audio files,
but bundle them into few large splits
(e.\,g.,  train, dev, test)
as in the case for Librispeech~\citep{panayotov2015librispeech}.\footnote{https://huggingface.co/datasets/librispeech\_asr}

\emph{Datasets} scales to very large datasets
as its data loading is based in Apache Arrow\footnote{https://github.com/apache/arrow}
and allows it to load datasets only partially into memory
and to stream datasets
when downloading them.
\emph{audb} always has to load whole tables into memory.
It offers two strategies for avoiding high memory consumption:
splitting into smaller tables and using partial loading
(see \cref{sec:partial-loading}).
An advantage of \emph{audb} is
that it uses pickled files,
which read faster than Apache Arrow files.

\emph{Datasets} does not support organising annotations
into different tables,
or referencing the same audio files
or parts of it multiple times.
Each data point that is returned contains
the actual audio signal,
a link to the corresponding audio file 
and associated labels.
Whereas in \emph{audb},  there is only a loose connection
between audio files and annotations.
This means there is exactly one copy of an audio file,
even when it is referenced from different tables
or different, possibly overlapping segmentation exist.
It further allows mapping annotations from one table to another.
For example, consider the following three tables:

\begin{lstlisting}
# ID: speakers
speaker,age
spk01,19
spk02,21
\end{lstlisting}

\begin{lstlisting}
# ID: files
file,speaker
a.wav,spk01
b.wav,spk02
\end{lstlisting}

\begin{lstlisting}
# ID: emotion
file,start,end,emotion
a.wav,0,0 days 00:00:01,happy
a.wav,0,0 days 00:00:02,calm
\end{lstlisting}

\noindent
If the `speakers' table is assigned as scheme
to the `speaker' column
of the `files' table,
its labels can be mapped to the values
of a column in the `speakers` table,
e.\,g.,  `age'.
And it can be requested using the segmentation
from the `emotion' table as index:

\begin{lstlisting}[language=Python]
db["files"]["speaker"].get(
    index=db["emotion"].index,
    map="age",
)
\end{lstlisting}

\noindent
The result is a segmented table with the age of the speakers:

\begin{lstlisting}
file,start,end,age
a.wav,0,0 days 00:00:01,19
a.wav,0,0 days 00:00:02,19
\end{lstlisting}


\section{Use Cases}

\subsection{Browsing and Searching Datasets}

\emph{audb} provides the possibility to list available datasets.

\begin{lstlisting}[language=Python]
datasets = audb.available(only_latest=True)
\end{lstlisting}

\noindent
The results can be filtered for datasets
that have a scheme `emotion':

\begin{lstlisting}[language=Python]
# Create scheme lookup dictionary
schemes = {}
for name, version in datasets.version.items():
    schemes[name] = list(
        audb.info.schemes(name, version=version)
    )
# Search for datasets with scheme "emotion"
emotional_datasets = [
    name for name in schemes
    if "emotion" in schemes[name]
]
\end{lstlisting}

\begin{figure}[t]
    \centering
    \includegraphics[width=\columnwidth]{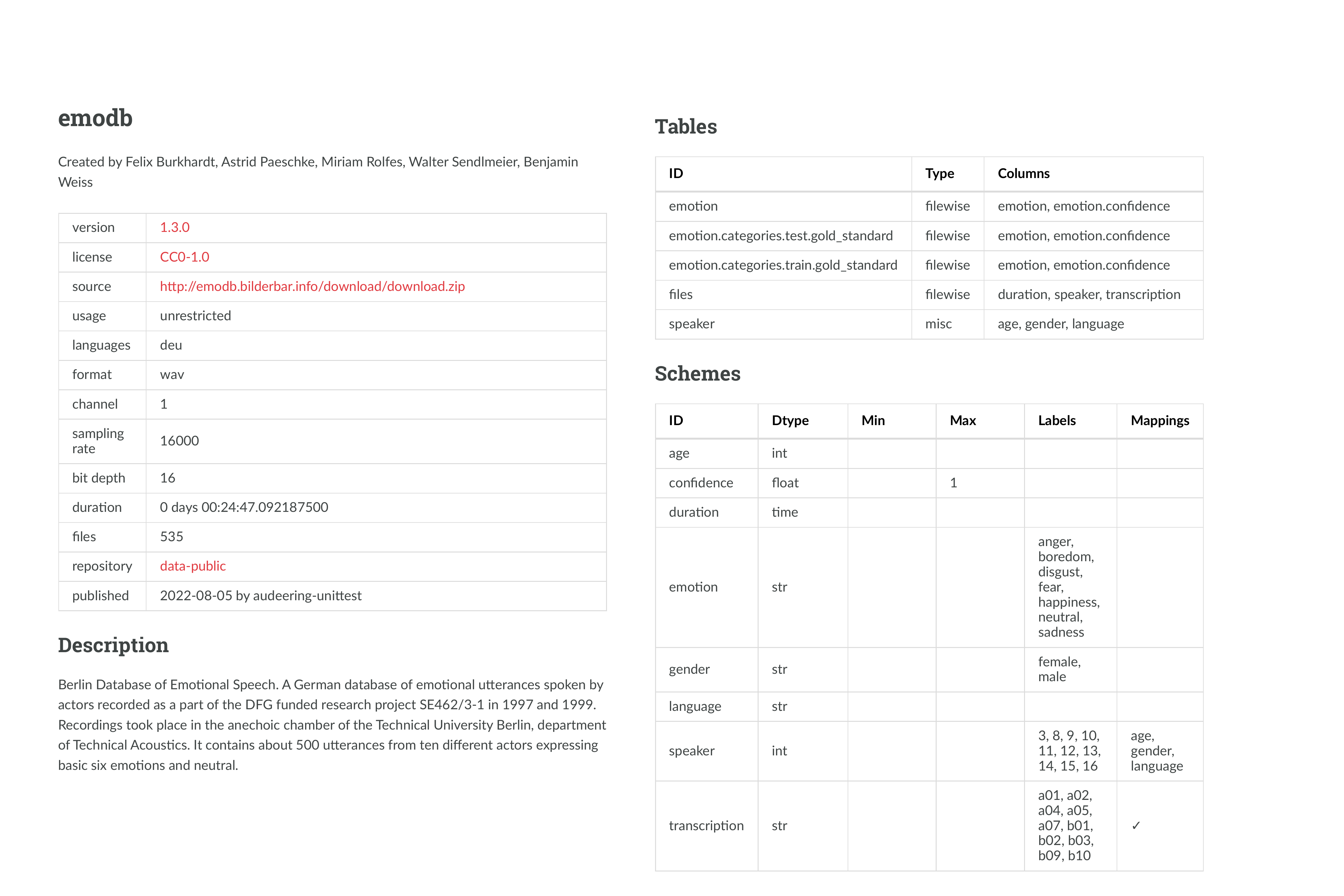}
    \caption{
    Excerpt of the data card for emodb~\citep{burkhardt2005emodb}.
    It includes a description of the dataset
    and metadata like author and license,
    and lists the tables, columns and schemes in the dataset.
    }
    \label{fig:emodb}
\end{figure}

\noindent
Since metadata and annotations are provided in a well defined format,
it is possible to automatically create documentation
in form of data cards or datasheets~\citep{gebru2021datasheets}.
\cref{fig:emodb} shows an example data card
for the emodb dataset~\citep{burkhardt2005emodb}.
It summarises the most important facts in a tabular form,
provides a long description of the dataset
together with an audio example
and lists available tables, columns, and schemes.
Data cards for all datasets available with the default public repositories
of \emph{audb}
are available at \url{https://audeering.github.io/datasets/}.


\subsection{Fine-tuning a Model for Emotion Recognition}

With the emergence of foundation models~\citep{bommasani2021opportunities}
pre-trained on large amounts of data,
it is nowadays a common task in the paralinguistic community
to fine-tune generic models to a specific problem.
The following example shows how this can be easily achieved
using \emph{audb} and \emph{audinterface}.

Assume we have a callable model
that converts an audio signal into a compact feature representation (embeddings).
We first create an interface for it:

\begin{lstlisting}[language=Python]
feature_extractor = audinterface.Process(
    process_func=model,
    num_workers=4,
)
\end{lstlisting}

\noindent
Then, we load a dataset and convert it into a single feature matrix
on which we train a linear model
that predicts the emotional content of the input signal:

\begin{lstlisting}[language=Python]
db = audb.load("emodb", version="1.3.0")
labels = db["emotion"]["emotion"].get()
features = feature_extractor.process_index(
    labels.index
)
\end{lstlisting}

\noindent
For a full example see \url{https://github.com/audeering/w2v2-how-to/blob/main/notebook.ipynb}.

\subsection{Publishing new dataset splits}

Datasets of emotional speech
might be published without an official train, dev, test split
like emodb~\citep{burkhardt2005emodb}
or IEMOCAP~\citep{busso2008iemocap}. 
Other datasets such as CommonVoice~\citep{ardila2020}
or VoxCeleb~\citep{nagrani2017voxceleb}
might miss splits for tasks not considered originally when collecting the data, e.\,g.,  age prediction.
\emph{audb} makes it easy to add new tables and splits to a dataset
and publish a new version,
encouraging other researchers to reuse them.
For example, with version 1.2.0 we added a train-test split
to the emodb dataset:

\begin{lstlisting}[language=Python]
audb.load_to("./db", "emodb", version="1.1.1")
# Add new splits to db
audb.publish("./db", "1.2.0", repository)
\end{lstlisting}

\section{Conclusion}

The recent success of foundation models~\citep{bommasani2021opportunities}
in the paralinguistic and audio community
has raised the need for a large number of diverse datasets
to train and evaluate models fine-tuned to a specific task~\citep{turian2022}.
\emph{audb} is a lightweight, yet powerful Python library
to publish, maintain and access
audio datasets and their annotations.
Its highlights are: a built-in versioning system, an automated workflow to publish and update datasets locally
or on a remote server,
and sharing datasets to specific end-users or communities.
We published a selection of publicly available datasets
in a public repository.
The repository is pre-configured in \emph{audb} and the datasets
can be directly accessed.
For a list of available datasets
please visit anonymised.
 
\section{Acknowledgements}

The authors would like to thank
Baha Eddine Abrougui,
Christian Geng,
Stephan Huber,
Andreas Triantafyllopoulos,
Damiano Zanardo
for their contributions,
and the pandas community for providing the basis for audb.
This research has been partly funded by the 
European EASIER 
(Grant Agreement number: 101016982) and by the European SHIFT 
project (Grant Agreement number: 101060660).

\section{\refname}
 
\printbibliography[heading=none]

\newpage

\end{document}